\documentclass[times,authoryear]{elsarticle}

\usepackage{jasr}
\usepackage{framed,multirow}

\usepackage{amssymb}
\usepackage{latexsym}
\usepackage[T1]{fontenc}

\usepackage{url}
\usepackage{xcolor}
\definecolor{newcolor}{rgb}{.8,.349,.1}

\usepackage[citebordercolor=white]{hyperref}
\usepackage{amsthm,amsmath}

\journal{Advances in Space Research}

\begin{document}

\verso{Tatiana V\'ybo\v{s}\v{t}okov\'a \textit{etal}}

\begin{frontmatter}

\title{Correlation of Anomaly Rates in the Slovak Electric Transmission Grid with Geomagnetic Activity}%

\author[1]{Tatiana \snm{V\'ybo\v{s}\v{t}okov\'a}}
\author[2,3]{Michal \snm{\v{S}vanda}\corref{cor1}}
\cortext[cor1]{Corresponding author: 
  Tel.: +420-605-577-166}
\ead{svanda@sirrah.troja.mff.cuni.cz}

\address[1]{Department of Surface and Plasma Science, Faculty of Mathematics and Physics, Charles University, CZ-182 00 Prague, Czech Republic}
\address[2]{Astronomical Institute, Faculty of Mathematics and Physics, Charles University, CZ-182 00 Prague, Czech Republic}
\address[3]{Astronomical Institute of the Czech Academy of Sciences, CZ-25165 Ond\v{r}ejov, Czech Republic}

\received{--}
\finalform{--}
\accepted{--}
\availableonline{--}
\communicated{--}

\begin{abstract}
The induction of electric currents in electric power distribution networks is a well-known effect of Earth-directed eruptive events. Inspired by recent studies showing that the rate of power-grid anomalies may increase after exposure to strong geomagnetically induced currents in mid-latitude countries in the middle of Europe, we decided to investigate such effects for Slovak electric-power distribution grid managed by the Slovak Electricity Transmission System (SEPS). We obtained a list of disturbances recorded in the maintenance logs by the SEPS company technicians with their dates and other details. 

Working with unadjusted data we found unexpectedly strong seasonal variations of failure rates. We show that this is a consequence of the strong dependence on meteorological conditions. In particular, mean temperature showed the strongest correlation with anomaly rate, while precipitation amount was just a little weaker in terms of the correlation. After the suppression of the meteorological dependence in the failure-rate series, we obtained a new temperature-independent dataset. Using this clean dataset we found an increase in failure rates in the periods of maxima of geomagnetic activity compared to the adjacent minima of activity.
\end{abstract}

\begin{keyword}
\KWD Geomagnetic activity \sep K index \sep Power distribution networks
\end{keyword}

\end{frontmatter}



\section*{Introduction}
The Sun is the closest star to the Earth and it is a dominant body of the Solar system. However, its influence is not limited only to the gravity force that controls the movement of all bodies in the Solar system, but it changes the properties of the environment which fills the whole interplanetary space. The uppermost layer of the solar atmosphere, the corona, is constantly expanding in the form of solar wind.

The Earth itself is not affected by the solar wind directly since its impact is largely shielded by the Earth’s magnetic field. The large-scale propagating structures in the interplanetary space that originate from the interaction of solar wind streams and transient solar eruptions cause disturbances in the geomagnetic field \citep{badruddin2016study} that pose a risk to the performance and reliability of both space-borne and ground-based technological systems and even endanger human life or health through radiation exposure.

Coronal mass ejections (CMEs) and their subset, magnetic clouds (MCs -- twisted magnetic flux tubes), carry a significant amount of magnetic flux, mass, and energy outward from the Sun to the interplanetary medium \citep{amari2003coronal}. In many cases, they are responsible for geomagnetic storms since they are usually associated with interplanetary shocks and long--duration southward interplanetary magnetic field (IMF), thus reconnection can occur on the dayside magnetopause \citep{borovsky2006differences}.

Magnetic reconnection is a key explosive phenomenon in collisionless plasma that converts magnetic energy to plasma kinetic energy through a change in the magnetic field topology. The injection of solar wind particles and changes in the magnetic field due to the reconnection generate a variety of currents in the magnetosphere-ionosphere system, such as cross-tail current, field-aligned currents, partial ring current, etc. A fraction of the tail current can be temporarily diverted through the ionosphere, allowing closure of the current wedge and causing perturbations in the auroral zone and at the middle latitudes \citep{mcpherron2017mid}. Sudden variations in the solar wind's dynamic pressure cause changes in the magnetopause and tail current systems.

These perturbations lead to geomagnetic storms and substorms during which energetic particles are injected into the inner magnetosphere along the magnetic field lines. The plasma sheet is convected inward and the current wedge may rise. During storms, the ring current (a westward current comprising ions drifting westward and electrons drifting eastward in heights roughly $\sim2-4$ Earth radii in the equatorial plane) is supplied with charged particles from the plasma sheet, leading to a significant and prolonged decrease of the geomagnetic field \citep{ganushkina2017space}. The decrease of the geomagnetic field strength can be measured via the Dst index.

During storms, there can also be strong horizontal currents in the ionosphere, which fluctuate in the ionosphere and can cause a time variation of the terrestrial magnetic field, which induces a voltage potential on the surface of the Earth \citep{koskinen2001space}. When such a potential encounters a closed circuit, it induces currents to flow in it. Power transmission grids are networks of long transmission lines and stations linked to the ground through grounded connections. Together with the ground, transmission lines form closed loops which, when being exposed to an electric potential, experience the induction of currents \citep{ebihara2021prediction}. Subsequently, disruption and damage in the system attached to the conductor may appear \citep{pirjola2000geomagnetically}.

The current required to produce the flux in the ferromagnetic core consists of the magnetization current (the current required to produce the flux in the core of the transformer) and the core-loss current (the current required to make up for the hysteresis and eddy current losses). The magnetisation current of efficient power transformers is very small (generally less than 1\% of the load current). The level of GIC depends on engineering characteristics of the transmission line but also on the overhead ionospheric currents and on the underlying conductivity structure. Thus, GIC magnitudes in electricity transmission lines can vary from tens to several hundred of amperes depending on the latitude and strength of the magnetospheric activity \citep{espinosa2019effects}. Compared to normal alternating currents (AC) that flow in the network, GICs are quasi-direct currents (DC) and even small values of GIC can have harmful impacts on transformers' operation \citep{girgis2012effects}.

Such small values of GICs can cause the transformers to saturate every half-cycle. 
Even the largest transformer takes only a few amperes of AC excitation current to energize its magnetic circuit under typical conditions. Magnetic core saturation in an exposed transformer can occur at GIC levels as low as 1--10 A, producing extremely large and highly distorted AC to be drawn from the power grid \citep{koen2003geomagnetically}. This amplified AC from saturation effects can pose risks to power networks directly due to increased reactive power demand that can cause voltage regulation problems.

Another threat of nearly equal concern arises from collateral impacts from highly distorted transformer currents, caused by the presence of GICs in the transformer. These distortions can cascade problems by disrupting the performance of other network structures, causing them to go offline. In addition, individual transformers may be damaged from overheating due to this unusual mode of operation, which can result in long-term outages to key transformers in the network.

Although high-latitude regions are more at risk from GICs, middle and low latitude regions may also be significantly impacted \citep{gaunt2007transformer,torta2012geomagnetically,zois2013solar,lotz2017extreme,tozzi2019preliminary,gil2021evaluating}. The first study of anomalies in the Czech power grid as a function of geomagnetic activity (defined by the K index computed from the measurements of the Earth’s magnetic field at a local magnetometer station Budkov) has already identified some statistically significant increases of the rate of anomalies around month-long periods of higher geomagnetic activity than around nearby periods of lower activity \citep{vybovst2019statistical}. Nevertheless, the relationship between geomagnetic events and anomalies remained opened in terms of causality. If we expect there is a causality between the higher rate of anomalies recorded during increased geomagnetic activity in the Czech Republic, we anticipate a similar behaviour of the Slovak distribution network.

\section*{Data}

In this study, we investigate the relationship between anomaly rates recorded in the Slovak distribution network and geomagnetic activity. Moreover, we compare them with our previous findings of the increased rate of disturbances recorded in the Czech power distribution network during the period of increased geomagnetic activity to point out the direct impact of GICs on the distribution networks even in the mid-latitude countries. 

\subsection*{The Electric Power Transmission Network in Slovakia}

Electricity transmission in Slovakia is provided by the only transmission system operator, which is Slovenská elektrizačná prenosová sústava, a.s. (hereinafter as SEPS). Electricity distribution in Slovakia is provided mainly by three operators of regional distribution systems -- Západoslovenská distribučná , a.s. (West regional distributor), Stredoslovenská distribučná, a.s. (Middle regional distributor), and Východoslovenská distribučná, a.s. (East regional distributor). In addition to these three largest distribution system operators, there are around 142 local distribution system operators in the country. A total of 2,138 km of 400 kV voltage level lines, 769 km of 220 kV lines, and 22 substations are operated by SEPS. It is a member of the international group of the European Network of the Transmission System Operators for Electricity ENTSO-E that actively supports market integration and cross-border cooperation with the neighbouring transmission systems. 

\begin{figure*}[!ht]
    \caption{Power system of the Slovak Republic}
    \centering
   \includegraphics[width=0.7\textwidth]{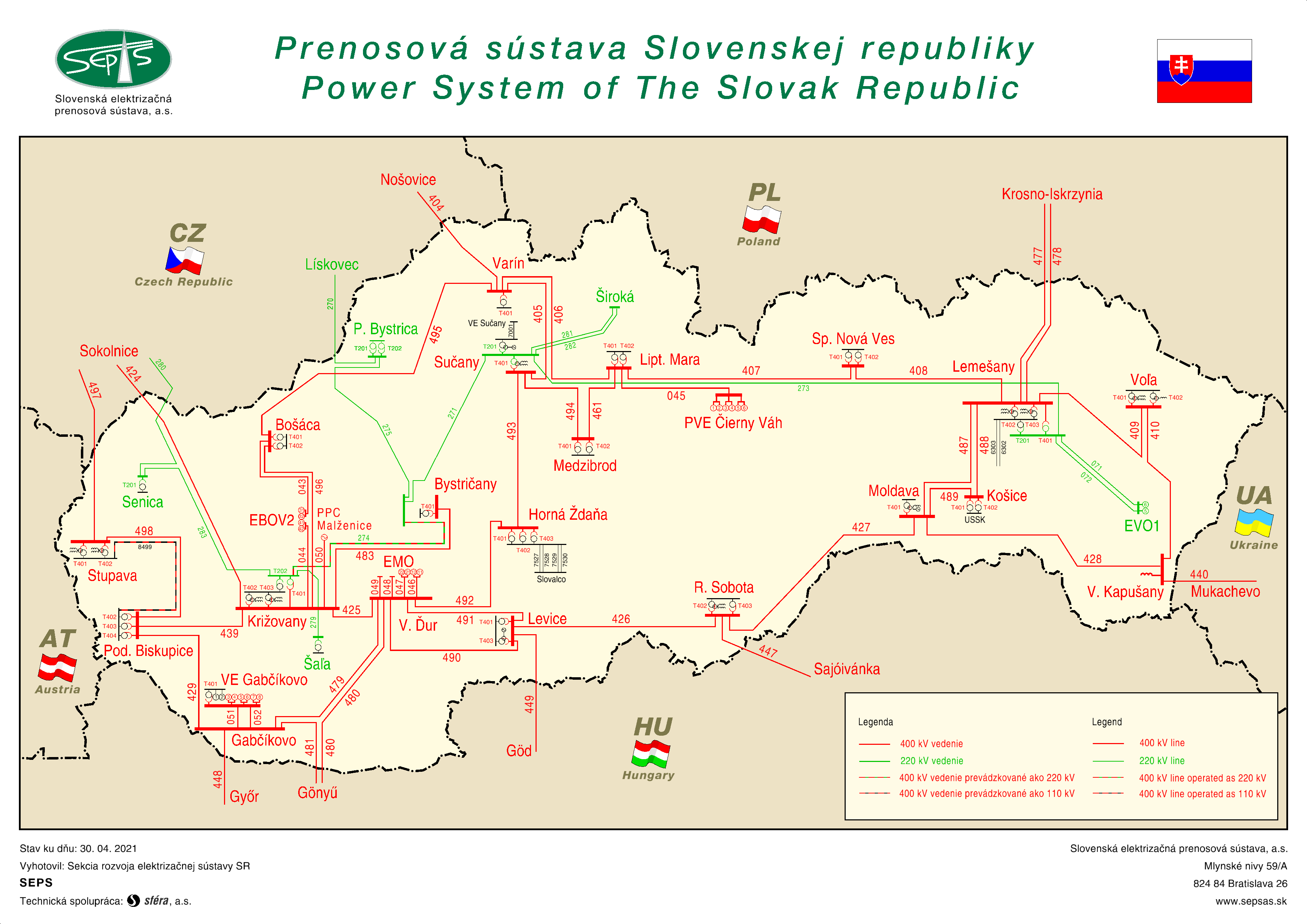}
    \label{fig:SEPS}
\end{figure*}

SEPS transmits electricity for the whole region of Slovakia and ensures the electricity transmission from power plants to the distribution networks and major customers connected to the $220$-kV and $400$-kV grids as can be seen in Fig.~\ref{fig:SEPS}. SEPS also provides import, export, and transit of electricity as well as exact measurements for transmission lines and substations. Almost $82 \%$ of the transmission system maintenance and attendance is ensured on a contractual basis by SEPS.

\begin{figure*}[!ht]
    \caption{Logs of Anomalies recorded by SEPS in comparison with K index}
    \centering
   \includegraphics[width=1.0\textwidth]{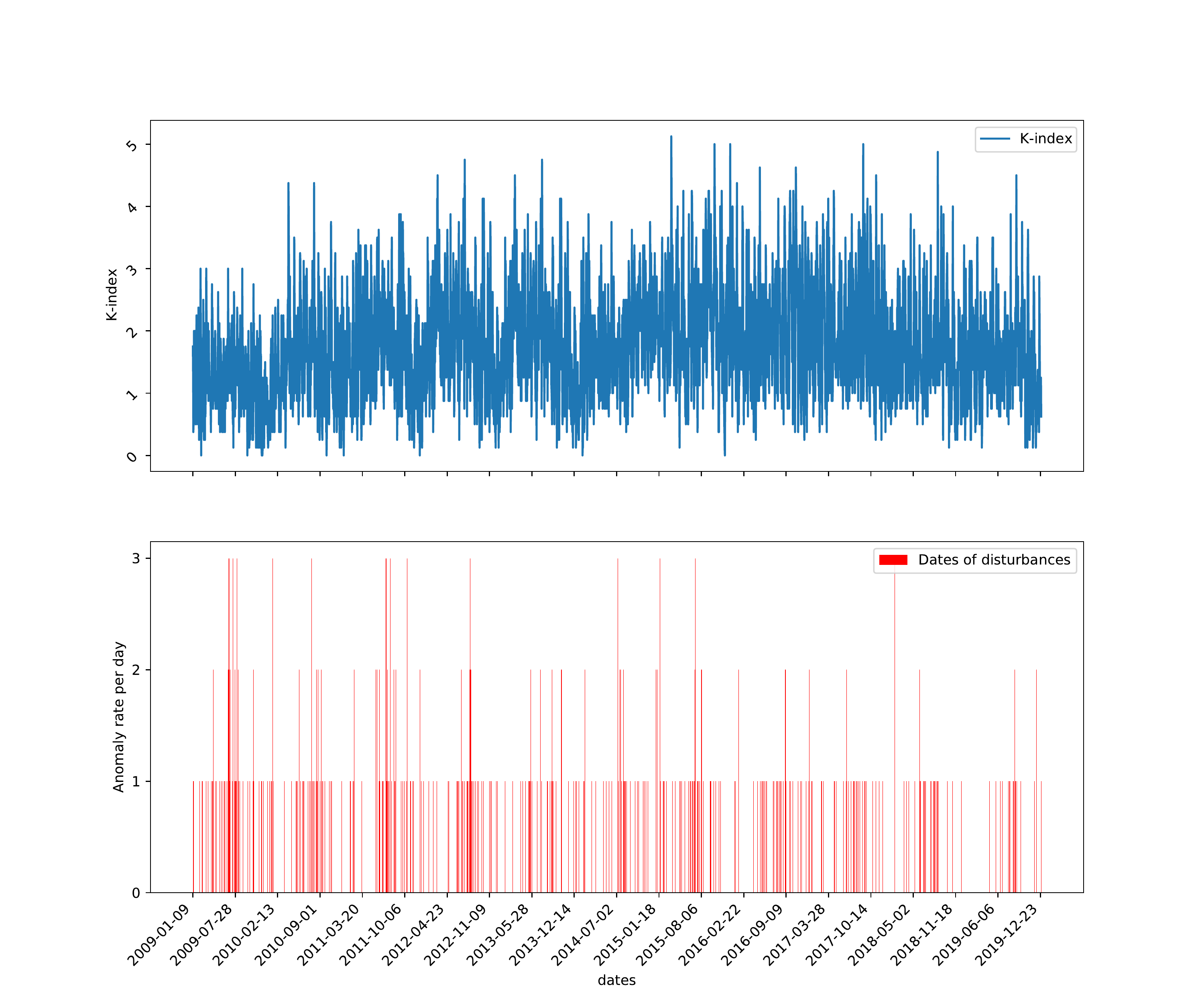}
    \label{fig:histogram}
\end{figure*}

\subsection*{Logs of Anomalies}

We obtained a list of recorded anomalies on the Slovak transmission grid maintained by SEPS company with detailed information about the disturbances (date, time, location, expected cause of the anomaly,\dots). For the purpose of our research, we work only with dates since our goal is to study the anomaly rates. The list includes not just equipment failure events (such as defects), but also events on the power lines, such as unscheduled switching, power outages, and service anomalies. For the purpose of this study, we removed the anomalies, where the indicated causes were obviously non-enviromental. Such removed items include for instance obvious human errors. The datasets are anonymized and must be provided as such due to a mutual non-disclosure agreement with SEPS. 

The total time span of recorded anomalies provided by SEPS is $10$ years ($2009$ -- $2019$). The final dataset is one by one compared with the level of geomagnetic activity using various statistical methods. The dataset is visualized in Fig.~\ref{fig:histogram} on the bottom panel.

\subsection*{Geomagnetic activity}
As discussed by \cite{schrijver2013disturbances}, choosing the right index to assess the effects of solar activity on the electricity distribution system is quite a difficult task. The problem with solar activity indices is that different eruptive events can have different geoeffectivity. In our study, we decided to use the local geomagnetic K index computed for Geomagnetic Observatory at Hurbanovo in Slovakia. Data are downloaded from the archive of K indices of Hurbanovo Observatory web page (http://www.geomag.sk/Archiv/). We chose the series computed using the Finnish Meteorological Institute method (FMI) and the threshold of 420~nT for ${\mathrm K}=9$. The steps undertaken for computation of K index using FMI method are as follows \citep{sucksdorff1991computer}:

\begin{enumerate}
    \item The raw magnetometer data is binned into average minute values. It is then cleaned with a moving hour long window removing spikes and obvious outliers. 
    \item For each 3 hour block the variation between the maximum and minimum of the two horizontal magnetic field values are compared to the Table.\ref{tab:kindex_FMI} to get an initial K-index. 
    \item For each hour of the day, the average horizontal values inside the interval centered on that hour with $(n + m)$ minutes on each side are calculated, where $n=K^{3.3}$ with $K$ being the initial estimate of the K-index, and $m$ is a constant which depends on the time of day as described by \cite{sucksdorff1991computer}. Together, they give an estimate of the solar-quiet variation.
    \item This estimate is then smoothed.
    \item The smoothed solar-quiet variation is subtracted from the raw data. This is then used as in step 2 to get a secondary K-index.
    \item Steps 3-5 are then repeated using the secondary K-Index to finally calculate the third and final K-index.
\end{enumerate}

\noindent We would like to point out that following the study by \cite{1995GeoJI.123..866M} the FMI methods best represents the K index of the geomagnetic field.

\begin{table}[]
    \centering
    \begin{tabular}{cccccc}
    \hline 
    K index value & Limit of Range Classes (nT) \\ 
    \hline 
        0&  0 -- 4.2 \\
        1&  4.2 -- 8.4 \\
        2&  8.4 -- 17 \\
        3&  17 -- 34 \\
        4&  34 -- 59 \\
        5&  59 -- 100 \\
        6&  100 -- 170 \\
        7&  170 -- 250 \\
        8&  250 -- 420 \\
        9&  $>$ 420 \\
    \end{tabular}
    \caption{Threshold limits for the K-index values used at Hurbanovo observatory. }
    \label{tab:kindex_FMI}
\end{table}

In order to compare values of the K index with the anomaly rate corresponding to each day, we computed the daily averages of 3-hour K indices. From the physical point of view, such value has a different meaning from the original quantity. However, we seek days or a period of increased/decreased geomagnetic activity, so we need to identify only local minima and maxima. Despite some interpretation concerns, the averaged K index series is a suitable quantity for this purpose.

\section*{Methodology}
Based on the analysis of disturbances in the Czech power grid where we looked for a link between anomaly rates and geomagnetic storms \citep{vybovst2019statistical} we repeat the process of data analysis. Using the same methods of processing, we ensure the correctness of the further comparison of results from the analysis of the Czech power grid with the current findings of susceptibility of the Slovak power grid to geomagnetic activity.

\begin{figure*}[!ht]
    \caption{Pearson correlation coefficient of original disturbance dataset with K index, mean temperature and precipitation amount for different smoothing windows. The line is plotted dotted in the case when the value of the correlation coefficient is not statistically significant (its $p$-value is lesser than 0.01). }
    \centering
    \setlength\abovecaptionskip{10pt}
   \includegraphics[width=1.0\textwidth]{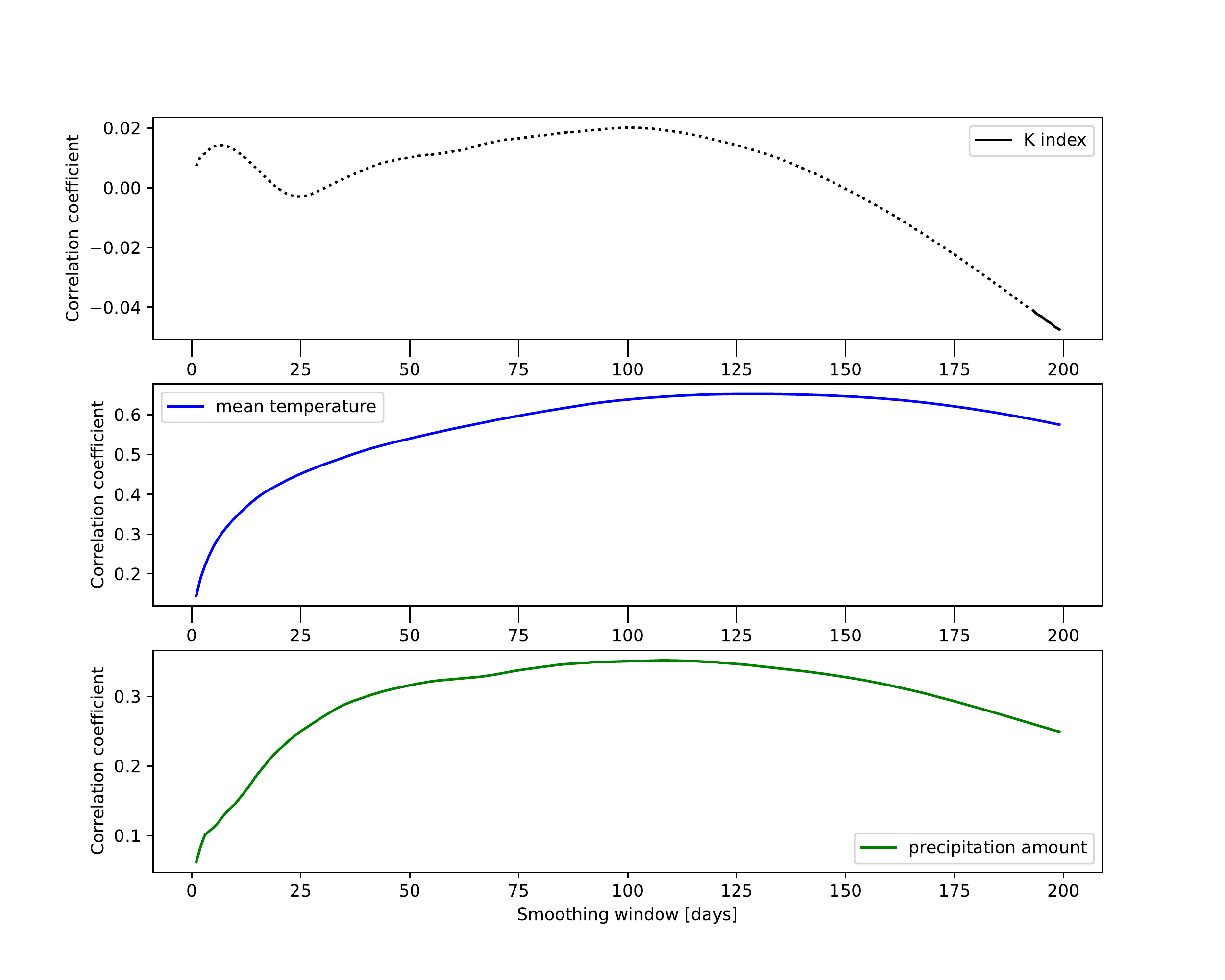}
    \label{fig:korelace_kindex_poruchy}
\end{figure*}

\begin{figure*}[!ht]
    \caption{Comparison of failure rate with K index, mean temperature and precipitation amount}
    \centering
    \setlength\abovecaptionskip{10pt}
   \includegraphics[width=1.0\textwidth]{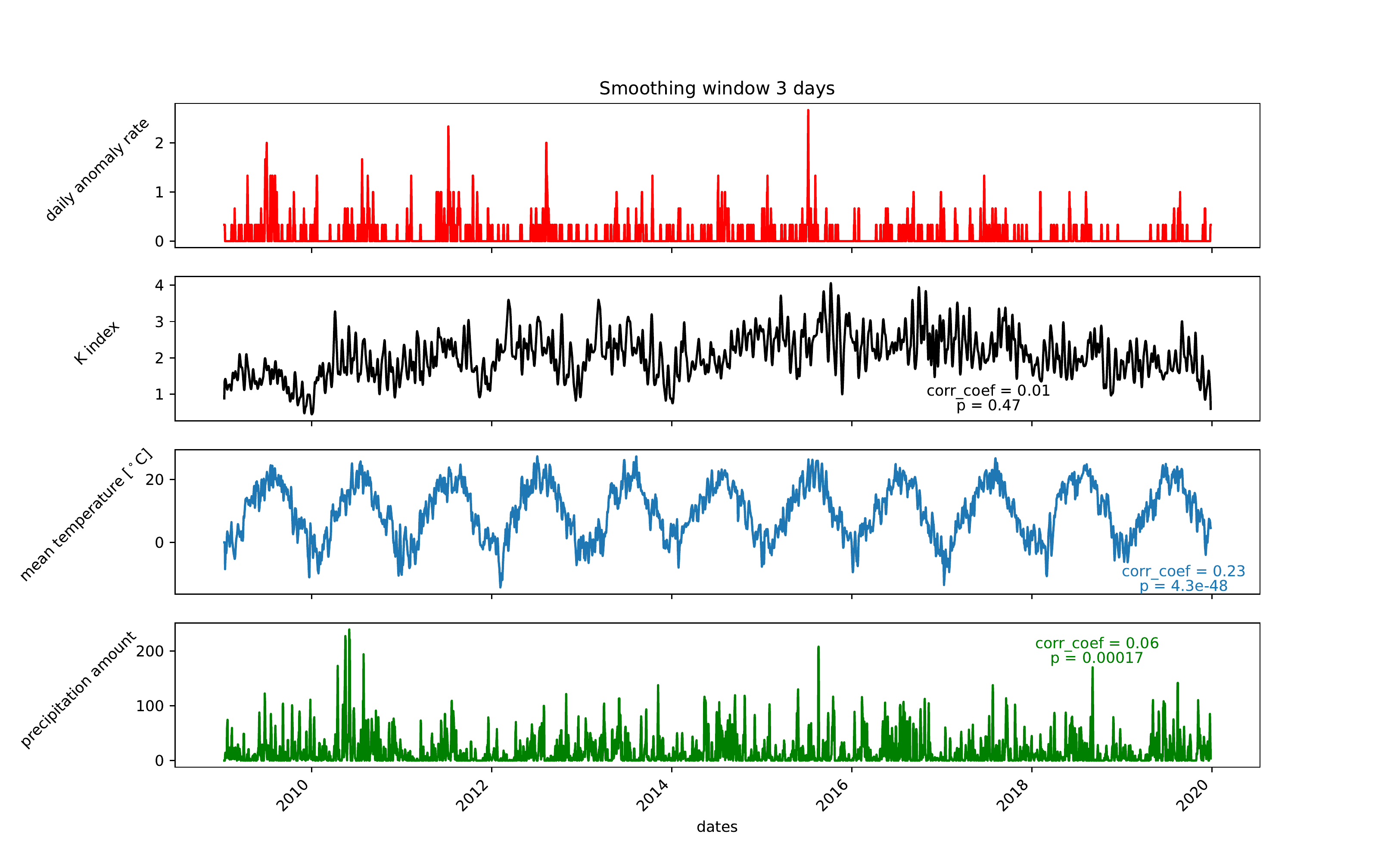}
\label{fig:kindex_poruchy}
\end{figure*}

The correlation coefficient of the anomaly rate with the K index is low for smoothing windows ranging from 10 to 200 days (see Fig.~\ref{fig:korelace_kindex_poruchy}, the upper panel). However, by eye, we noticed a strong apparent seasonal periodicity in the anomaly rates (as depicted, for instance, in the upper panel of Fig.~\ref{fig:kindex_poruchy}). We did not notice this behaviour when analysing equivalent data series for the Czech operators, hence we investigated the possible dependence of the anomaly rates on meteorological measurements (mean daily temperature and daily precipitation amount), which naturally contain quasi-periodic seasonal variations. 

\begin{figure}[!ht]
    \caption{Dependence of daily failure rate on mean temperature. Green points are the actual data points (the 3-day smoothed values of daily anomaly rates and mean temperatures as shown by the red and blue plots in Fig.~\ref{fig:kindex_poruchy}), black point with error bars indicate the normal points in 1-$^\circ$C bins (mean of daily anomaly rates in each 1 degree Celsius bin). The size of the error bar on each normal point is determined as the standard deviation of daily anomaly rates in each 1 degree Celsius bin. The red line represents the polynomial fit to the actual data points. }
    \centering
   \includegraphics[width=0.5\textwidth]{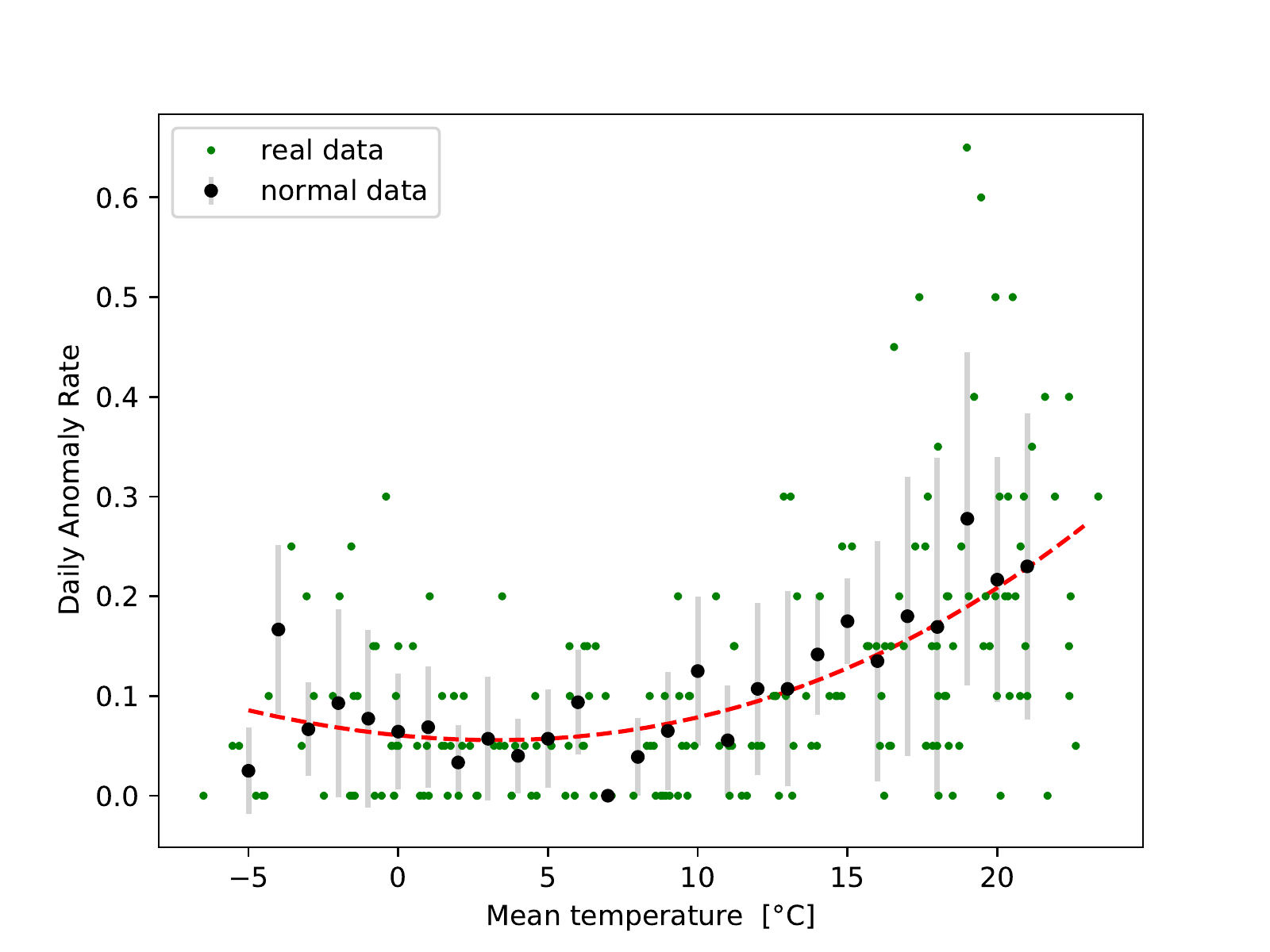}
    \label{fig:fit}
\end{figure}

We downloaded meteorological data from \mbox{http://kycera.eu/archivpocasia.php} website and computed their mean values from three separated stations: Ko\v{s}ice, Poprad-Tatry, and Hurbanovo. These data series somehow represent the average quantities over the Slovak territory, as the anomaly logs are also aggregated from the whole country. The data series averaged over the three indicated stations are plotted in the lower panels in Fig.~\ref{fig:kindex_poruchy} together with the corresponding correlation coefficients with the anomaly rates. 

We found that the mean temperature has the strongest correlation with the anomaly rate (see Fig.~\ref{fig:korelace_kindex_poruchy}, where, e.g., the correlation coefficient has a value of $0.4222$ for the smoothing window of 20 days and is statistically significant, its $p$-value is practically zero). Given that the greatest correlation is with temperature, it is suspected that temperature plays the most important role. Therefore, a simplified model was sought to describe this dependence. This is done by plotting the binned and smoothed temperature and the number of disturbances against each other, where it turned out that it could be solidly fit with a polynomial, see Fig.~\ref{fig:fit}. Our simplified model has a straightforward interpretation. The devices operate optimally within technical specifications at a certain temperature, whereas both the increase and decrease of the outside temperature may lead to issues (e.g., cooling issues in the case of increased temperature) resulting in the occurrence of anomalies. The power lines are affected by temperature changes, too. 

When plotting the anomaly rates as a function of the mean temperature, the polynomial fit represents the dependence satisfactorily: 

\begin{equation}
    {\mathrm{DAR}} = a(T-T_0)^3 + b(T-T_0)^2 + c(T-T_0),
\label{eq:fit}
\end{equation}
where $a$,$b$,$c$ and $T_0$ are coefficients of the polynomial fit determined by {\sc Python} using the $curve\_fit$ function from {\sc Scipy} package, $T$ corresponds to the smoothed mean temperature. In our case, $T_0 \approx 3^\circ$C represents the minimum of the temperature function. In other words, it may be interpreted as the temperature at which the operation of the power grid is optimal. To suppress the noisy behaviour, we smoothed and binned both the temperature and the daily-averaged rate series with three-day windows. The length is determined from the empirical experience that in central Europe, the weather changes typically every three days. The quantity ${\mathrm{DAR}}$ then represents the binned and smoothed daily anomaly rate.

\begin{figure*}[!ht]
    \caption{Comparison of original, residual and modelled datasets. At the top of the figure, we plot original failure logs and also their smoothed value over 3 days. The yellow curve corresponds to modelled disturbances while the residuum (original - modelled disturbances) and its smoothed version are plotted in blue. To complete the graph with full information we add a smoothed K index, mean temperature and precipitation amount at the bottom of the figure }
    \centering
    \setlength\abovecaptionskip{10pt}
   \includegraphics[width=1.0\textwidth]{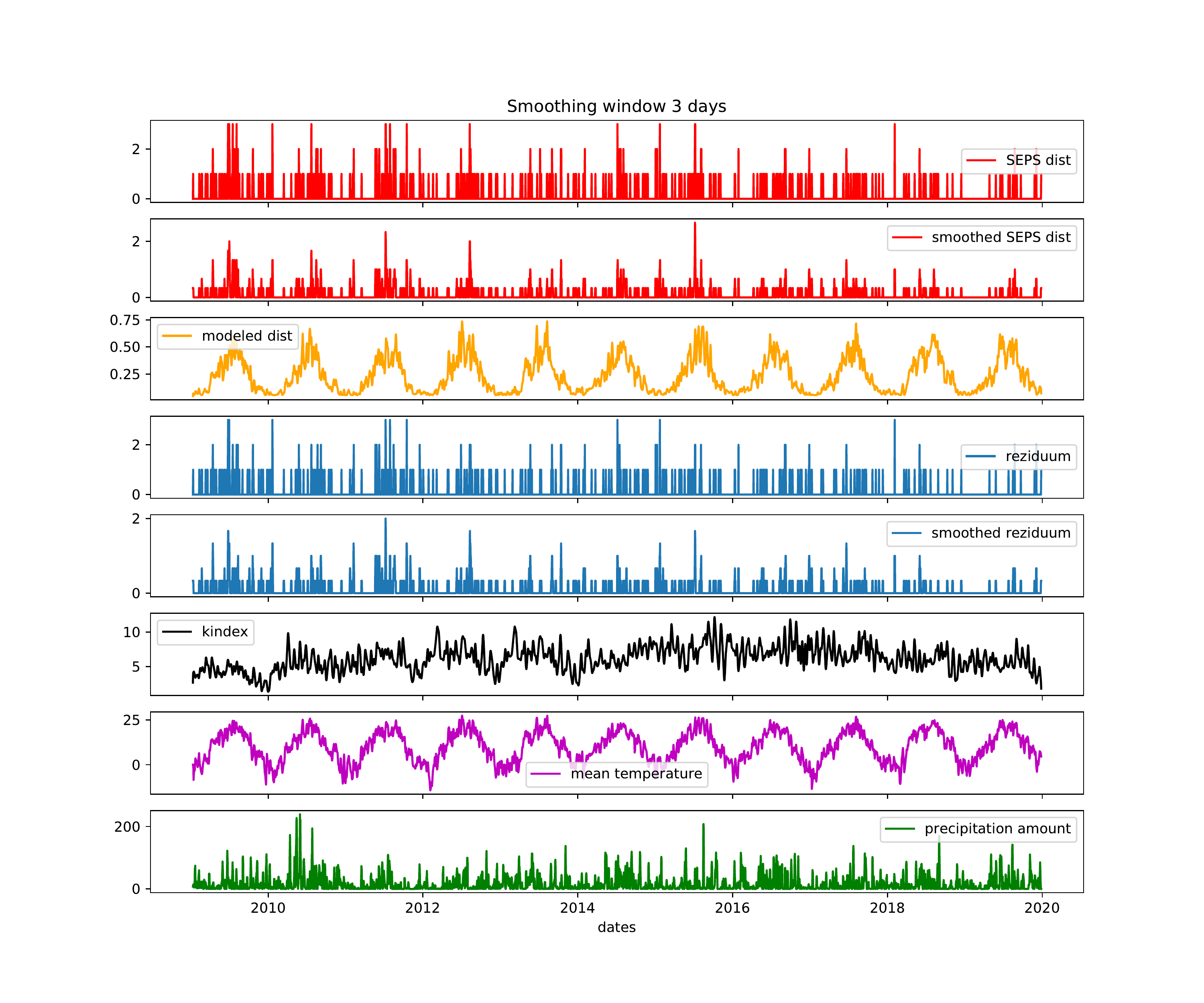}
    \label{fig:modelled_data}
\end{figure*}

\begin{figure*}[!ht]
    \caption{Correlation coefficient of residual disturbances with K index, mean temperature and precipitation amount for different smoothing windows. The dotted line indicates the correlation coefficients which are not statistically significant. }
    \centering
    \setlength\abovecaptionskip{10pt}
   \includegraphics[width=1.0\textwidth]{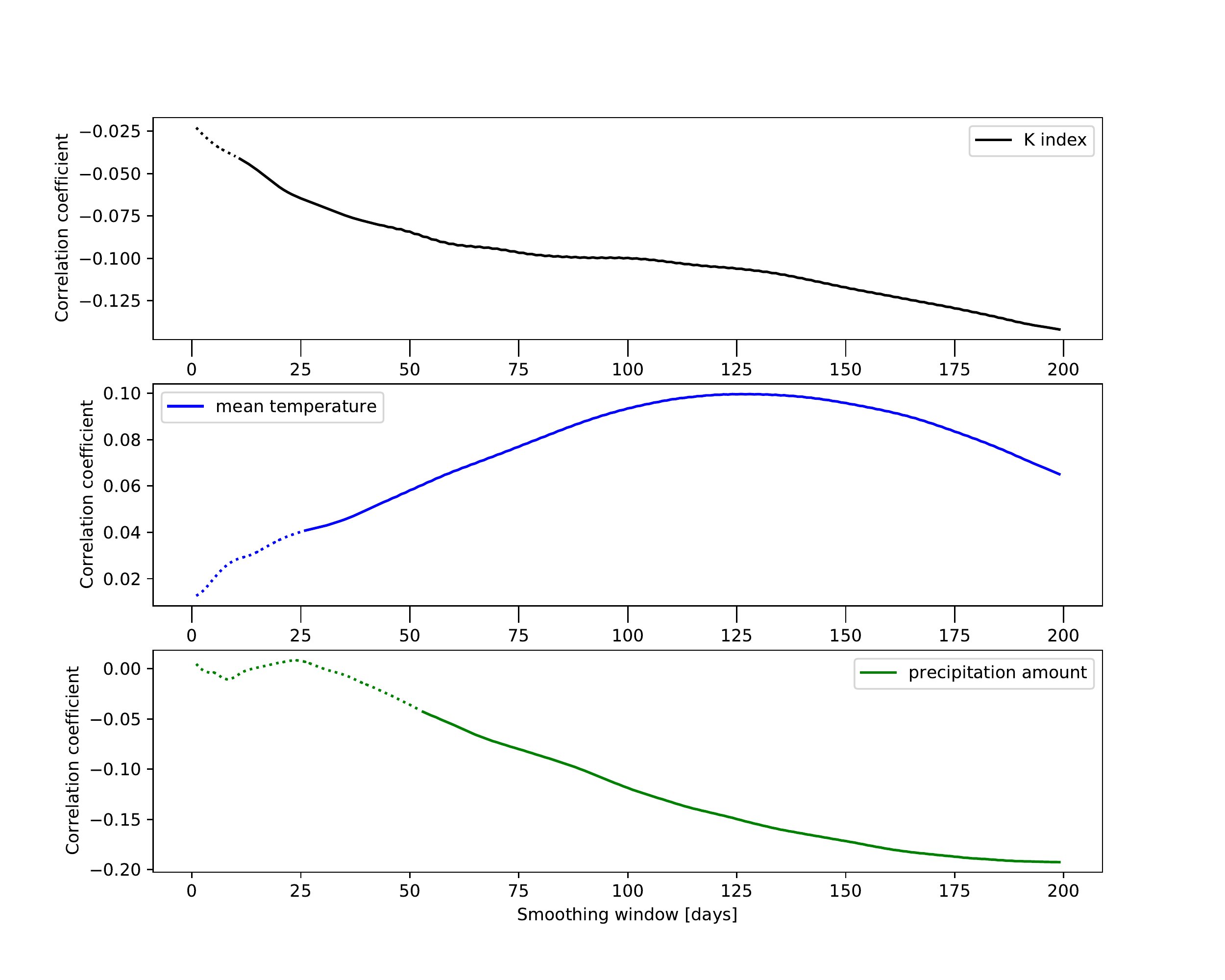}
    \label{fig:modelled_correlation}
\end{figure*}

Using the equation (\ref{eq:fit}) for each day we modelled the expected ${\mathrm{DAR}}$s when taking only the temperature dependence into account (see the third panel in Fig.~\ref{fig:modelled_data}). This temperature-dependent model was subtracted from the actual daily anomaly rates. In this way, we obtained the residuum (the fourth panel in Fig.~\ref{fig:modelled_data}), which was further investigated by the below given statistical methods. We tested that the model effectively removed the correlation of the residuum with the mean temperature, as demonstrated in Fig.~\ref{fig:modelled_correlation} in the middle panel. The correlation coefficient of the residuum with the mean temperature is less than 0.1 for all considered accumulation windows. 

A new corrected series of failure rates (the residuum) is used as a new input to our code for the statistical analysis of disturbances and geomagnetic activity. Our methodology is consistent with the retrospective cohort study with tightly matched controls \citep[for overview, see][]{Masao_Iwagami202222005}. For convenience reasons, we are testing the alternative hypothesis. The alternative hypothesis states that there is no difference between the number of anomalies registered in the periods around local maxima of activity and around local minima of activity. The hypothesis testing is performed by using the binomial test. 

The Slovak Republic is a rather small country, also the total length of the transmission lines is rather small. This leads to a series of daily anomaly rates, which is rather scarce, reaching maximally 3 registered anomalies in one day (see Fig.~\ref{fig:histogram}). With such counts, it is impossible to perform a meaningful statistical analysis directly. Moreover, studies by other authors pointed out that the occurrence of anomalies may be delayed with respect to the exposure to GICs \citep{zois2013solar}. Therefore, we integrated the anomalies in the accumulation windows having the width $W$. The analysis is done for the set of windows $W$, where the lengths range from 10 to 100 days. We chose such interval simply because smaller windows increase the noise levels meaning that lesser numbers of disturbances fall into the tested intervals. On the other hand too large values of $W$ cause the K index series to be overly smoothed. The accumulation window of 100 days may seem way too long. \cite{koen2003geomagnetically} investigated the effects of GICs in the South-African network and performed also the dissolved-gas analysis of the transformers' oil bash. They found out that the deterioration continues after the initial damage caused by GICs and may lead to failure after a long time. The delay between the exposure and the failure also depends on the transformer loading and possibly also on other stresses. The authors also pointed out that the damage caused by repeated exposure to low-level GICs cumulates and may finally manifest even after the ``last'' low level event.  

For comparability, the reference series (the K index) is smoothed with the same window $W$ to make sure that the variability of both series is comparable. For smoothing we're using two custom made function in Python. The average temperature values from the previous days ($W$) are considered for the given day in case of $smooth\_previous$ which we're using for smoothing of mean temperature and precipitation amount. The reason is that previous, not future, events are expected to have an impact, so classic (two-sided) smoothing may not be representative. To smooth K index and anomaly rate we are using $boxcarsmooth$ function that first creates a smoothing window in length of window $W$ and uses it in convolution with actual input which is K index or disturbances.

\begin{figure}[!ht]
    \caption{Our code recognize local minima and maxima in the daily averaged K index (blue solid curve). Bars represent intervals with a width of $W=70$ days around these extrema. The code looks for non-overlapping, closest-in-time pairs of maximum (blue) and minimum (red) }
    \centering
   \includegraphics[width=0.5\textwidth]{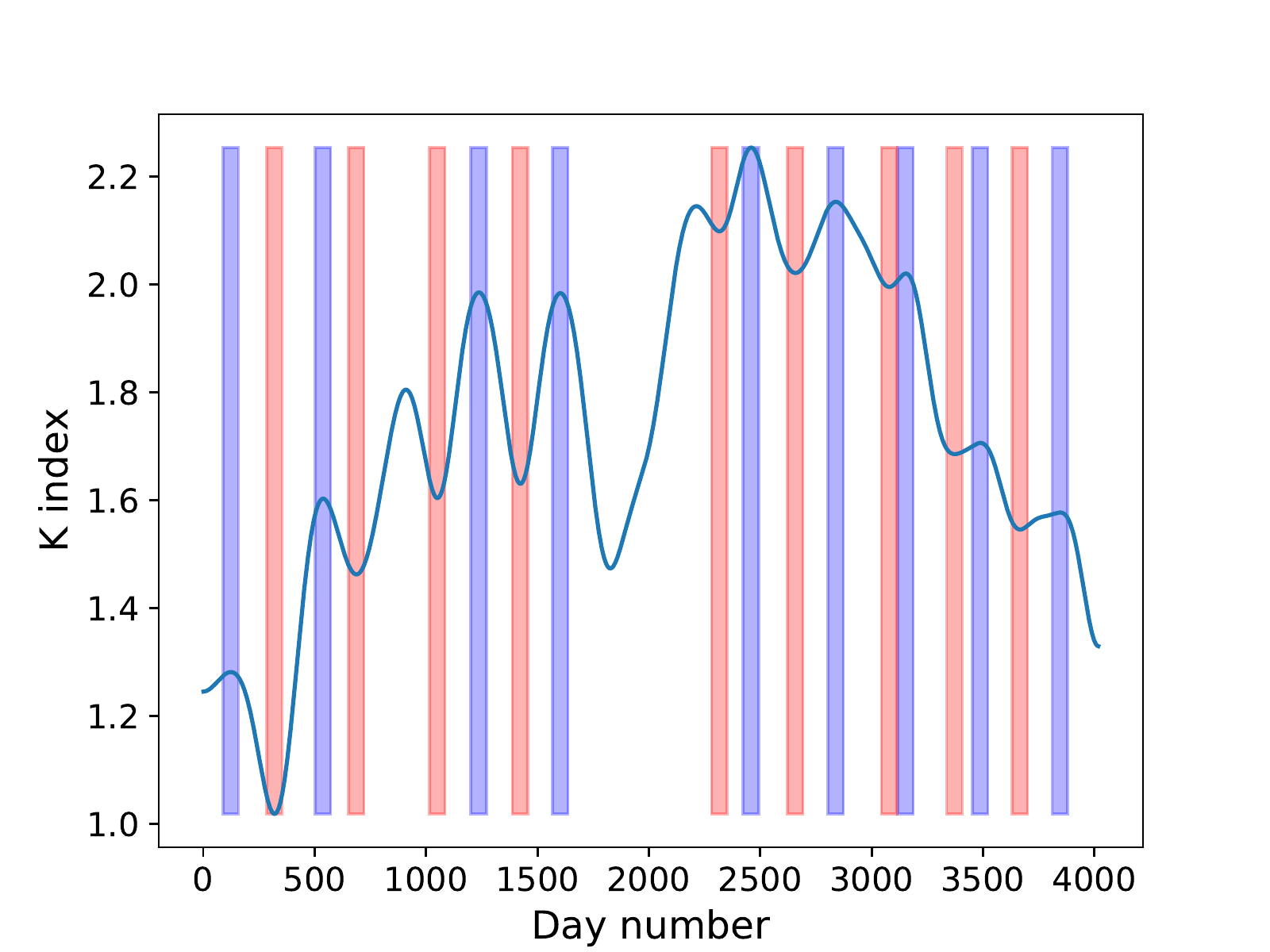}
    \label{fig:plot_70}
\end{figure}

Our analysis code written in {\sc Python} works as follows. In the reference series, we identify the local maxima and minima in geomagnetic activity. Around each maximum and each minimum, a window with the length $W$ is placed. The locations of the maxima are positioned at the beginnings of such intervals, while the locations of the minima are placed in the centres of the intervals. The code performs the pairing of the windows around the minima and maxima (to ensure tightly matched controls). We use several criteria while selecting the maximum/minimum pair. First of all, the individual $W$-days long intervals cannot overlap. At the same time, they need to be paired close in time, so we can minimize possible secular trends, represented, for instance, by the development of the grid, by climate changes, or by secular loading changes. Moreover, a pair of selected maxima and minima of geomagnetic activity needs to be unique within the whole dataset. An example of the local extrema detection and their pairing is given in Fig.~\ref{fig:plot_70}. The maxima and minima not fulfilling the requirements are downselected. 

Then, the {\sc Python} code computes the total number of anomalies within the remaining intervals of maxima in geomagnetic activity, $N_{\mathrm H}$, as well as the sum of anomalies during the remaining geomagnetic activity minima intervals, $N_{\mathrm L}$. 
According to our initial hypothesis, we would expect that during the maxima of the geomagnetic activity there will be a larger total number of anomalies than during the activity minima:

\begin{equation}
    N_{\mathrm H} > N_{\mathrm L}.
\end{equation}

The statistical significance of the differences is tested by using a binomial test \citep{binom_wiki}. The binomial test states the probability $P$ that the differences between the total count of anomalies registered in the low geomagnetic activity windows $N_{\mathrm L}$ and in the high geomagnetic activity windows $N_{\mathrm H}$ are in accordance with the model (with the alternative hypothesis). If $P$ is lower than 5$\%$ (our selection of statistical significance), then the alternative hypothesis may be rejected. In this situation, we found that the difference in counts $N_{\mathrm L}$ and $N_{\mathrm H}$ is statistically significant and cannot be only due to the statistical realisation of these numbers. $P$ is computed as

\begin{equation}
P_{\rm H,L}=2\sum_{k=x}^{n} {\binom{n}{k}} p^k(1-p)^{n-k},
\label{eq:probability}    
\end{equation}
where $n = N_{\mathrm H} + N_{\mathrm L}$. The parameter $p$ states the model-expected probability of the disturbance occurring during the high-activity intervals. In the tested (that is in the alternative) hypothesis, we assume that the probability of the disturbance occurring during the maximum or the minimum will be the same, i.e. $p = 1/2$. Finally, $x={\mathrm{max}}(N_{\mathrm H}, N_{\mathrm L})$. We used the binomial test to determine whether or not the disturbance rate is influenced by geomagnetic activity. Based on the $P$-value, we can say if we can detect the difference in the anomaly rates occurring during the minima and maxima of geomagnetic activity for the specified dataset and accumulation window $W$. 

We have also the meteorological data available, so we took advantage of having the working code at our disposal and performed a similar statistical analysis when taking the temperature and precipitation amount as the reference series instead of the K index.

Furthermore, we compute the relative risk $R$.
In the statistical analysis of cohort studies, relative risk is used to determine the strength of the association between exposure and outcomes. It makes use of two samples: one that was exposed to a specific causal characteristic (geomagnetic activity) and another that was not. If the output coming from the computation of relative risk $R$ is $1$ there is no difference between the two tested groups. In the case of $R > 1$, there are more positive cases in a group with a causal attribute (exposure to geomagnetic activity) and vice versa. We calculated the number of days with disturbances and the number of days without disturbances for both the intervals around the local maxima and minima. The relative risk was then calculated as follows:

\begin{equation}
    R = \frac{\alpha}{\alpha+\beta} / \frac{\gamma}{\gamma+\delta}, 
\end{equation}
where $\alpha$ denotes the number of days with anomalies and $\beta$ denotes the number of days without anomalies, both for intervals of increased solar activity, $\gamma$ is the number of days with an anomaly, and $\delta$ is the number of days without one for intervals with lower solar activity.

\section*{Results}

\begin{table}[!ht]
    \caption{Statistical analysis of all SEPS recorded anomalies with K index}
    \centering
    \begin{tabular}{cccccccccc}
    \hline 
    W & Intervals & $N_H$ & $N_L$ & $P_{HL}$ & $R$ \\ 
\hline 
10 & 52 & 52 & 51 & {1.000} & 0.96\\
20 & 19 & 40 & 32 & 0.410 & 1.25\\
30 & 15 & 52 & 49 & 0.842 & 1.07\\
40 & 11 & 41 & 41 & 1.000 & 0.93\\
50 & 10 & 41 & 40 & 1.000 & 1.07\\
60 & 9 & 63 & 39 & 0.022 & 1.55\\
70 & 9 & 77 & 46 & 0.007 & 1.67\\
80 & 7 & 88 & 39 & $<$10$^{-3}$ & 2.13\\
90 & 5 & 74 & 37 & $<$10$^{-3}$ & 1.83\\
100 & 4 & 68 & 36 & 0.002 & 1.72\\

\hline
    \end{tabular}
    \label{tab:kindex}
\end{table}

\begin{table}[!ht]
    \caption{Statistical analysis of residual SEPS anomalies with K index}
    \centering
    \begin{tabular}{cccccccccc}
    \hline 
    W & Intervals & $N_H$ & $N_L$ & $P_{HL}$ & $R$ \\ 
\hline 
10 & 52 & 50 & 41 & 0.402 & 1.21\\
20 & 19 & 34 & 31 & 0.804 & 1.22\\
30 & 15 & 42 & 51 & 0.407 & 0.92\\
40 & 11 & 38 & 44 & 0.581 & 0.94\\
50 & 10 & 37 & 41 & 0.734 & 1.05\\
60 & 9 & 57 & 41 & 0.129 & 1.24\\
70 & 9 & 64 & 49 & 0.188 & 1.17\\
80 & 7 & 78 & 42 & 0.001 & 1.49\\
90 & 5 & 63 & 40 & 0.030 & 1.30\\
100 & 4 & 61 & 39 & 0.035 & 1.26\\

\hline
    \end{tabular}
    \label{tab:kindex_modeled}
\end{table}

\begin{table}[!ht]
    \caption{Statistical analysis of all SEPS recorded anomalies with mean temperature}
    \centering
    \begin{tabular}{cccccccccc}
    \hline 
    W & Intervals & $N_H$ & $N_L$ & $P_{HL}$ & $R$ \\ 
\hline 
10&  22 & 43 & 32 & 0.248 & 1.26\\
20&  8 & 46 & 18 & 0.001 & 2.76\\
30&  7 & 60 & 23 & $<$10$^{-3}$ & 2.82\\
40&  8 & 78 & 34 & $<$10$^{-3}$ & 2.71\\
50&  8 & 92 & 40 & $<$10$^{-3}$ & 2.50\\
60&  8 & 101 & 46 & $<$10$^{-3}$ & 2.44\\
70&  7 & 95 & 42 & $<$10$^{-3}$ & 2.52\\
80&  6 & 90 & 32 & $<$10$^{-3}$ & 2.96\\
90&  6 & 96 & 32 & $<$10$^{-3}$ & 3.20\\
100&  7 & 125 & 38 & $<$10$^{-3}$ & 3.23\\

 \hline
    \end{tabular}

    \label{tab:MT}
\end{table}

\begin{table}[!ht]
    \caption{Statistical analysis of residual SEPS anomalies with mean temperature}
    \centering
    \begin{tabular}{cccccccccc}
    \hline 
    W & Intervals & $N_H$  & $N_L$ & $P_{HL}$  & $R$ \\ 
\hline 
10&  22 & 6 & 9 & 0.607 &  0.64\\
20&  8 & 7 & 4 & 0.549 & 1.06\\
30&  7 & 9 & 6 & 0.607 &  0.87\\
40&  8 & 12 & 7 & 0.359 &  0.88\\
50&  8 & 13 & 9 & 0.523 & 0.88\\
60&  8 & 14 & 10 & 0.541 &  0.88\\
70&  7 & 15 & 9 & 0.307 &  0.92\\
80&  6 & 14 & 5 & 0.064 &  1.15\\
90&  6 & 15 & 6 & 0.078 &  1.17\\
100&  7 & 25 & 6 & 0.001 & 1.68\\

\hline
    \end{tabular}
    \label{tab:MT_modeled}
\end{table}

\begin{table}[!ht]
    \caption{Statistical analysis of all SEPS recorded anomalies with precipitation amount}
    \centering
    \begin{tabular}{cccccccccc}
    \hline 
    W  & Intervals & $N_H$  & $N_L$ & $P_{HL}$ & $R$ \\ 
\hline 
10&  44 & 53  & 30 & 0.015 &  1.43\\
20&  23 & 71  & 58 & 0.291 &  1.17\\
30&  13 & 68  & 38 & 0.005 &  1.90\\
40&  10 & 94  & 38 & $<$10$^{-3}$ &  2.31\\
50&  9 & 108  & 35 & $<$10$^{-3}$ &  2.93\\
60&  8 & 102  & 38 & $<$10$^{-3}$ &  2.56\\
70&  8 & 103  & 40 & $<$10$^{-3}$ &  2.34\\
80&  7 & 103  & 39 & $<$10$^{-3}$ &  2.45\\
90&  6 & 71   & 30 & $<$10$^{-3}$ &  2.44\\
100&  5 & 69  & 36 & 0.002 &  2.00\\

\hline
 
    \end{tabular}
    \label{tab:PA}
\end{table}

\begin{table}[!ht]
    \caption{Statistical analysis of residual SEPS anomalies with precipitation amount}
    \centering
    \begin{tabular}{cccccccccc}
    \hline 
    W  & Intervals & $N_H$ & $N_L$ & $P_{HL}$  & $R$ \\ 
\hline 
10&  44 & 11 &  7 & 0.481 &  1.32\\
20&  23 & 17 &  10 & 0.248 &  1.14\\
30&  13 & 17 &  8 & 0.108 &  1.30\\
40&  10 & 20 & 8 & 0.036 &  1.19\\
50&  9 & 17 &  8 & 0.108 &  1.02\\
60&  8 & 14 &  7 & 0.189 &  1.05\\
70&  8 & 16 &  7 & 0.093 &  1.12\\
80&  7 & 18 &  7 & 0.043 &  1.05\\
90&  6 & 10 &  4 & 0.180 &  1.58\\
100&  5 & 11  & 6 & 0.332 &  1.22\\

\hline
    \end{tabular}
    \label{tab:PA_modeled}
\end{table}

We were able to investigate the effects of various parameters in the code on the results using our {\sc Python} code. For each run, the software required two main inputs: a series of daily-averaged K-index values and a series of dates of anomalies occurring in the electric power transmission network. Similarly, we used the same code to test the relationship of daily anomaly rate with mean temperature or precipitation amount. We study the sensitivity of the failure rate to these external conditions while using the same methodology.

One can easily notice that for smaller windows $W$, the number of disturbances falling into this interval is smaller compared to the number of disturbances in larger windows $W$. We expect such behaviour since the smaller the window $W$, the smaller smoothing of the K index which leaves us with shallower local extrema. On the other hand, in the case of large smoothing, the peaks are over-smoothed and pairs of maxima/minima can be hundreds of days apart. 

For the original anomaly-rate series, we found that for all three datasets compared (K index, mean temperature, precipitation amount), there always is a larger or equal number of disturbances during the period of larger K index (Table \ref{tab:kindex}), during the period of increased mean temperature (Table \ref{tab:MT}) or the period of increased precipitation amount (Table \ref{tab:PA}). At the same time, there is a strong statistical significance of the difference between the number of failures during increased periods for $W \ge 60$ days. Moreover, the relative risk is always larger than $1$ with the largest value of $2.13$ for the K index and $W=80$ days. Such a large correlation between the K index and the anomaly rates is not expected. Very small $P$-values for the temperature as the reference series confirms that the anomaly rates have a significant dependence on temperature, which is (accidentally) also correlated with the K index due to the semi-annual variations of the geomagnetic field \citep[see e.g.][]{1852RSPT..142..103S,1959RSPTA.251..525M,2002JASTP..64...47C}. The statistical analysis strengthens the need to remove the temperature-related trend from the anomaly rates and to focus on the residuum after the removal of this trend. 

We repeat the same process for the disturbance residuum. In the case of the K index, we still have more disturbances during the period of increased geomagnetic activity than during the quiet period except for windows of 30, 40 and 50 days. In those cases, the differences are negligible and probably caused by the small amount of disturbances falling into smaller windows while for larger windows we still have strong indications that the increase in the number of failures for geomagnetic active days is not just an accidental observation for windows with more than $80$ days based on the $P$-value. We can see such behaviour probably because of the larger number of recorded disturbances in the power grid when considering the large accumulation windows. As expected, the relative risk $R$ is larger than $1$ for these windows. That is also demonstrated in an example Table \ref{tab:kindex_modeled}. Same behaviour was observed in statistical analysis of disturbances recorded in the Czech distribution grid.

Comparison of corrected failure logs and mean temperature showed a higher number of disturbances during days of increased mean temperature than during days with lower temperature, however, a binomial test proved that such difference is not statistically significant except for the largest window with 100 days as can be seen in Table \ref{tab:MT_modeled}. This behaviour confirms that our model satisfactorily suppresses the temperature-dependent trend in the anomaly rates. Similarly, we found no statistical significance in the difference between the number of failures during the period of increased precipitation amount than during the quiet period, see Table \ref{tab:PA_modeled}. We can say that we removed dependency on meteorological conditions from the original failure log provided by SEPS and found that for such dataset there is still a strong correlation with the number of disturbances during geomagnetic active days as it was also shown for disturbances recorded in the Czech power distribution network. Value of relative risk $R$ is always larger than $1$ for precipitation amount while in case of mean temperature analysis $R > 1$ holds only for windows larger or equal to $80$ days and $W = 30$. The $R$ value is meaningless when the number of recorded disturbances is low due to the statistical insignificance of $R$. On the other hand, when the number of disturbances is large, then $R = 1$ by definition. 




\section*{Concluding Remarks}
Our goal was to confirm the statistical analysis of the correlation between anomalies in the Czech electric power grid and the geomagnetic activity described by the K index, but this time with anomalies recorded in the Slovak electric power grid. Using the binomial test and relative risk methods of statistical analysis, we compared the failure rates in tens of days long intervals around local maxima of geomagnetic activity to periods surrounding local minima of geomagnetic activity.

We showed that the number of disturbances recorded by SEPS depicted a statistically significant increase during the local maxima of the geomagnetic activity, represented by the K index determined for Hurbanovo Geomagnetic Observatory.  We identified a strong dependence of the anomaly rates on the mean temperature and a weak correlation with the precipitation amount. This correlation is positive for all lengths of the accumulation window. We needed to suppress this strong meteorological dependence in the series of anomaly rates in order to be able to investigate the possible effects of the geomagnetic activity only. We showed that our simplified model suppresses the temperature dependence in the residuum series satisfactorily. 

When working with the residuum series, we confirm the statistical increase of grid anomaly rates during the local maxima of the geomagnetic activity compared to the adjacent minima. The increase of the anomaly rates is in particular significant for the accumulation windows larger than 70 days. Similar behaviour was discovered in the case of the dataset of disturbances recorded on the transmission and high-voltage distribution networks in the Czech Republic. 

The relative increase is roughly between 10 and 15 per cent. Furthermore, the percentage increase of the anomaly rate against a basal (determined from the geomagnetically quiet days) failure rate is around 11\%.

All findings mentioned above lead us to the conclusion that both Slovak and Czech power distribution grids are mildly susceptible to increased solar and geomagnetic activity, respectively, even though each dataset of failure logs is maintained by individual and independent company technicians. Since we found a hint that the disturbances indeed are caused by effects of the geomagnetic activity in the Czech Republic, we expect to find similar behaviour for the neighbouring countries with relatively similar geography. Our study confirms similar studies performed recently in the neighbouring Czech Republic \citep{vybovst2019statistical} and Poland \citep{gil2020transmission}. 

\section{Acknowledgments}
The results presented in this paper rely on data collected at magnetic observatories. We thank Geomagnetic Observatory Hurbanovo that supports magnetic measurements for the region of Slovakia. Magnetic data may be downloaded from their webpage  www.geomag.sk/Archiv. Log of anomalies was provided by SEPS. We are grateful to data providers for giving us an opportunity to exploit their logs of anomalies, namely, to \v{L}. Holka, P. Výbo\v{s}\v{t}ok  and Z. Kody\v{s}. The maintenance logs are considered strictly private by the power company and are provided under nondisclosure agreements. The meteorological data were downloaded from \mbox{http://kycera.eu/archivpocasia.php}. We also thank both referees for their comments that helped to improve the manuscript. ASU CAS is funded by the institute research project ASU:67985815. Authors' contributions: MŠ and TV designed the research which was then performed by TV within her doctoral research. TV negotiated the provision of the data with SEPS, drafted the manuscript, and plotted the figures. Both authors contributed to the final manuscript.

\bibliographystyle{jasr-model5-names}
\biboptions{authoryear}
\bibliography{biblio}      

\end{document}